\documentclass{PoS}
\pdfoutput=1

\usepackage{amsmath}
\usepackage{cite}

\newcommand{\ie}{{\it i.e.}}

\newcommand{\eg}{{\it e.g.}}

\newcommand{\eq}{Eq.}

\newcommand{\Ref}{Ref.}
\newcommand{\Refs}{Refs.}

\newcommand{\equ}[1]{\eq~(\ref{equ:#1})}

\newcommand{\figu}[1]{Fig.~\ref{fig:#1}}

\title{Multi-messenger interpretation of the neutrinos from TXS 0506+056}

\ShortTitle{Multi-messenger interpretation of the neutrinos from TXS 0506+056}

\author{\speaker{Walter Winter}$^1$\thanks{
This work has been supported by the European Research Council (ERC) under the European Union's Horizon 2020 research and innovation programme (Grant No. 646623).} , 
Shan Gao$^1$, Xavier Rodrigues$^1$, Anatoli Fedynitch$^2$, \newline Andrea Palladino$^1$, Martin Pohl$^{1,3}$
        \\
        $^1$Deutsches Elektronen-Synchrotron (DESY), Platanenallee 6, D-15738 Zeuthen, Germany\\
        $^2$Department of Physics, University of Alberta, Edmonton, Alberta, Canada T6G 2E1\\
        $^3$Institute of Physics and Astronomy, University of Potsdam, 14476 Potsdam, Germany\\
        E-mail: \email{walter.winter@desy.de}$^*$}

\abstract{
We discuss possible interpretations of the neutrinos 
observed from the AGN blazar TXS 0506+056 in the multi-messenger and 
multi-wavelength context, including both the 2014-15 and 2017 neutrino 
flares. While the neutrino observed in September 2017 has to describe 
contemporary data in \eg\ the X-ray and VHE gamma-ray ranges, data at 
the 2014-15 excess are much sparser. We demonstrate that in both cases 
the simplest possible one-zone
AGN blazar models face challenges. While the 2017 flare can be well 
interpreted by considering more sophisticated source geometries, the 
2014-15 flare is much harder to describe with conventional models. One challenge is the energy injected into the electromagnetic cascade coming together with the neutrino production, which cannot be reconciled with the 13 observed neutrino events. We also speculate if a common interpretation of both flares is feasible.}

\FullConference{36th International Cosmic Ray Conference -ICRC2019-\\
		July 24th - August 1st, 2019\\
		Madison, WI, U.S.A.}

\begin{document}

\section{Introduction}

An electromagnetic follow-up campaign of the neutrino event IceCube-170922A indicated that this event came from the direction of a known Active Galactic Nucleus (AGN) blazar named TXS 0506+056~\cite{TXS_MM} -- which was at that time flaring at multiple wavelengths; this is likely to be the first discovery of an astrophysical neutrino beyond TeV energies from a known object. A further analysis of archival IceCube data revealed that this blazar was emitting neutrinos before: within the period between October 2014 and March 2015 an excess of $13 \pm 5$ events over background~\cite{TXS_orphanflare} was found. Archival electromagnetic data during that period were, however, sparse, and only available in the gamma-ray~\cite{Aartsen:2019gxs}, as well as radio and optical ranges~\cite{Padovani:2018acg}. Moreover, there was no significant electromagnetic flaring activity, except for a possible indication of a  gamma-ray hardening at GeV energies~\cite{Padovani:2018acg,Aartsen:2019gxs}.

Theoretical descriptions of these events include neutrino production models via $p \gamma$ interactions~\cite{Gao:2018mnu,Cerruti:2018tmc,Zhang:2018xrr,Keivani:2018rnh,Ahnen:2018mvi,Gokus:2018lgx,Righi:2018xjr,Halzen:2018iak,Rodrigues:2018tku,Reimer:2018vvw} -- which are frequently invoked in lepto-hadronic models to describe the spectral energy distribution (SED) of AGN blazars -- and $pp$ interactions~\cite{Liu:2018utd,Wang:2018zln}. In both cases, primary cosmic rays (the protons) need to be accelerated to energies in the PeV range, which may then interact with radiation or gas. While $pp$ interactions necessarily require more freedom (or parameters) to describe the gas target, $p \gamma$ interactions may occur with the radiation which can self-consistently describe the SED; we therefore focus on $p \gamma$ models here. These models typically describe the first hump in the characteristic two-hump SED of AGN blazars with synchrotron radiation off acceleracted electrons, whereas the second hump comes from inverse Compton scattering in leptonic models, or from various processes induced by co-accelerated protons in lepto-hadronic models (such as proton synchrotron radiation, gamma-rays from $\pi^0$ decays, or synchrotron/inverse Compton scattering off secondary electrons/positrons; see \eg\ \cite{Gao:2016uld} for a more detailed discussion). 

A theoretical interpretation of the 2014-15 and 2017 neutrino flares should predict the right flux and event rate. Since only one neutrino event was observed in 2017, there is substantial uncertainty on the expected neutrino flux from the source, and there are good reasons to expect that the predicted neutrino event rate should be much smaller. In short, if many similarly neutrino-faint objects existed, as expected from gamma-ray population studies, many such coincidences would be expected if each source really produces one neutrino event on average~\cite{Palladino:2018lov,Strotjohann:2018ufz}.  Another uncertainty comes from a difference between the effective area of the alert system and that of the throughgoing muon analysis (triggering the alert system requires a substantially higher flux because of quality cuts). From these considerations, the theoretical model probably needs to predict between about 0.01 and 0.1 events per source for the 2017 period. The situation is very different for the 2014-15 neutrino flare, for which archival muon track data detected with the corresponding muon neutrino effective area were used. Because of the high statistics (13 events) the predicted neutrino event number needs to be of a similar magnitude. Note, however, that a power law assumption for the extraction of this event number was used in the analysis, and it is unclear if the event rate extraction changes for different assumptions on the spectral shape.

In this work, we focus on the main challenges and the main conclusions from theoretical models used to describe the 2014-15 and 2017 neutrino flares which can self-consistently describe the electromagnetic SED as well. Examples include the relevance of X-ray and VHE gamma-ray data as direct hadronic signatures, the conclusions which can be obtained from the time response of the SED during the 2017 electromagnetic flare, and the missing electromagnetic activity during the 2014 neutrino flare. 

Our method is solving a coupled partial differential equation system for all relevant species (electrons, photons, protons, and secondaries), following \Ref~\cite{Gao:2016uld}, and performing extensive parameter space scans with millions of model computations (counted without analytical re-scalings). The neutrino production rate will be proportional to the proton density times radiation density. The proton density scales with proton injection luminosity (related to the gamma-ray luminosity by the baryonic loading) and the confinement time of the protons. The radiation density depends on the observed photon luminosity and the geometry/size of the source, described by a spherical blob with radius $R'$ in the jet rest frame and moving with a relativistic Doppler factor $\Gamma$. Further parameters of the model include the description of the injected electron and proton spectra (power laws with certain minimal and maximal energies), the (known) redshift of the source, and possible parameters describing external radiation fields, if applicable, see \Refs~\cite{Rodrigues:2017fmu,Gao:2016uld,Gao:2018mnu,Rodrigues:2018tku,2019NatAs...3...24P} for details.

\section{One neutrino associated to a gamma-ray flare in 2017}

Let us first of all focus on the neutrino event IceCube-170922A associated to the flaring state of TXS 0506+056. The simplest possible modeling of the SED from AGN blazars are 
so-called one-zone models, where the radiation is emitted from a radiation zone  which is typically assumed to be spherical in the jet frame -- apart from an acceleration zone from which the particles are injected; see \Refs~\cite{Gao:2018mnu,Cerruti:2018tmc,Keivani:2018rnh,Gokus:2018lgx,Sahakyan:2018voh}. Although a purely leptonic (synchroton self-Compton, SSC) model can, in principle, describe the SED~\cite{Gao:2018mnu,Cerruti:2018tmc} (for a more critical perspective, see \cite{Keivani:2018rnh}), such a model will not produce any neutrinos. A popular lepto-hadronic alternative postulates that the second hump of the SED comes from $\pi^0$ decay photons~\cite{Mannheim:1993jg}, which can be ruled out for TXS 0506+056 because it would lead to an excess in X-rays by orders of magnitude~\cite{Gao:2018mnu}. 

X-ray and very-high energy (VHE) gamma-ray data have in fact been found to be crucial for constraining the hadronic contribution, although the latter may be absorbed in the extragalactic background light (EBL) for high-redshift sources~\cite{Gao:2018mnu,Ahnen:2018mvi,Keivani:2018rnh,Murase:2018iyl}. There are two reasons for that: on the one hand, the signatures of secondary gamma-rays and leptons produced by the proton interactions will, either directly or after re-processing, be  dominant in these energy ranges -- as the measured fluxes are relatively low (lower than at the optical and gamma-ray peaks). On the other hand, protons interact with target photons of energy
\begin{equation}
 E_\gamma\, [\mathrm{keV}] \simeq  \frac{0.01 \, \Gamma^2}{E_\nu \, [\mathrm{PeV}]} \, 
 \label{equ:target}
\end{equation}
to produce neutrinos, where $\Gamma$ is the Doppler factor of the production region moving relativistically towards us. This means that X-ray fluxes determine the neutrino production efficiency at PeV energies. 

\begin{figure}[t]
\begin{center}
\includegraphics[width=6.5cm]{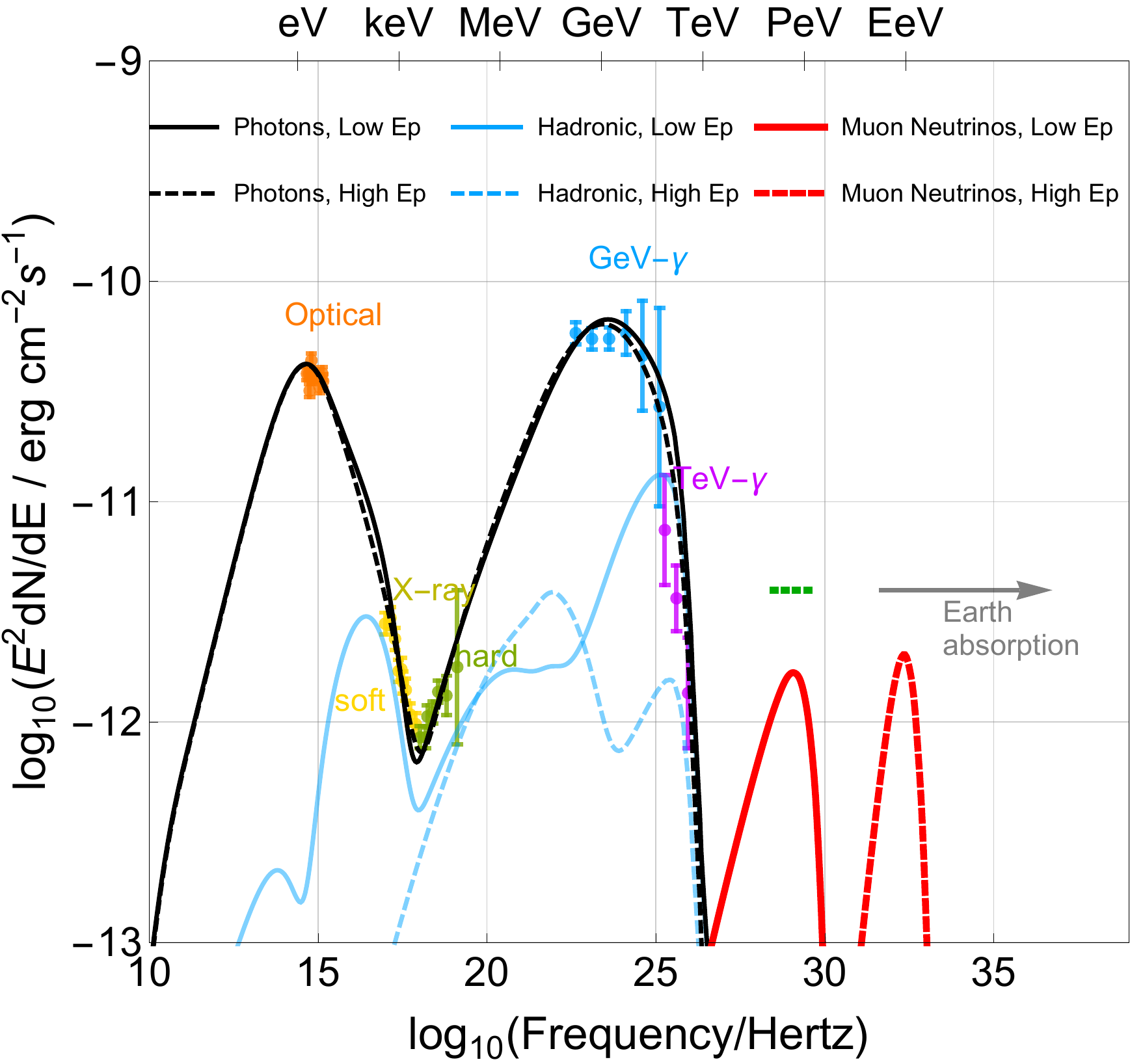}
\end{center}
\caption{\label{fig:OneZone} SED (black curves) and expected neutrino flux (red curves) as a function of frequency/energy for two versions of a one zone model during the electromagnetic flare corresponding to IceCube-170922A: $E_{\rm p, max} \sim 4.5$ PeV (solid curves) and $E_{\rm p, max} \sim 1700$ PeV (dashed curves). Hadronic contributions are shown in blue. Figure taken from \Ref~\cite{Gao:2018mnu} (Supplementary Materials). }
\end{figure}

Consequently, X-ray data limit the maximal possible neutrino flux, as it is illustrated in \figu{OneZone} for a conventional one-zone model with two different maximal proton energies. While this model describes the SED very well, the hadronic contribution (blue) is sub-dominant and the 
neutrino event rate is limited to about 0.1-0.2 during the flare (in consistency with the arguments used earlier). The expected neutrino energy depends on the maximal proton energy assumed, and the figure illustrates that a connection to ultra-high energy cosmic rays implies a neutrino spectrum peaking at too high energies (dashed curves).

Although being relatively simple in terms of the number of parameters, all studied one-zone models have limitations. A prominent method to estimate the physical luminosity available for the jet is the so-called Eddington luminosity estimating the accretion from a disk in the steady limit. Because of the low neutrino production efficiency, most one-zone models exceed the Eddington luminosity by far. A possible alternative is to increase the maximal proton energy, as in the dashed model in \figu{OneZone}, which leads to higher neutrino energies from lower target photon energies according to \equ{target}. From the SED in \figu{OneZone}, one can easily see that the photon spectrum strongly increases below keV energies (with decreasing energy), which means that the neutrino production efficiency will increase correspondingly, and that the energetics problem can be alleviated -- at the expense of a neutrino peak energy not matching observations. In summary, one-zone models may describe data if one either accepts that the Eddington luminosity is substantially (by several orders of magnitude) exceeded during the flare, or that the neutrino was measured at an energy significantly below the peak of the neutrino spectrum. Note that model alternatives, such as the proton synchroton model (for which the second hump comes from synchrotron radiation off the protons) face similar challenges~\cite{Cerruti:2018tmc,Gao:2018mnu,Keivani:2018rnh}.

\begin{figure}[t]
\begin{center}
\includegraphics[width=.53\columnwidth]{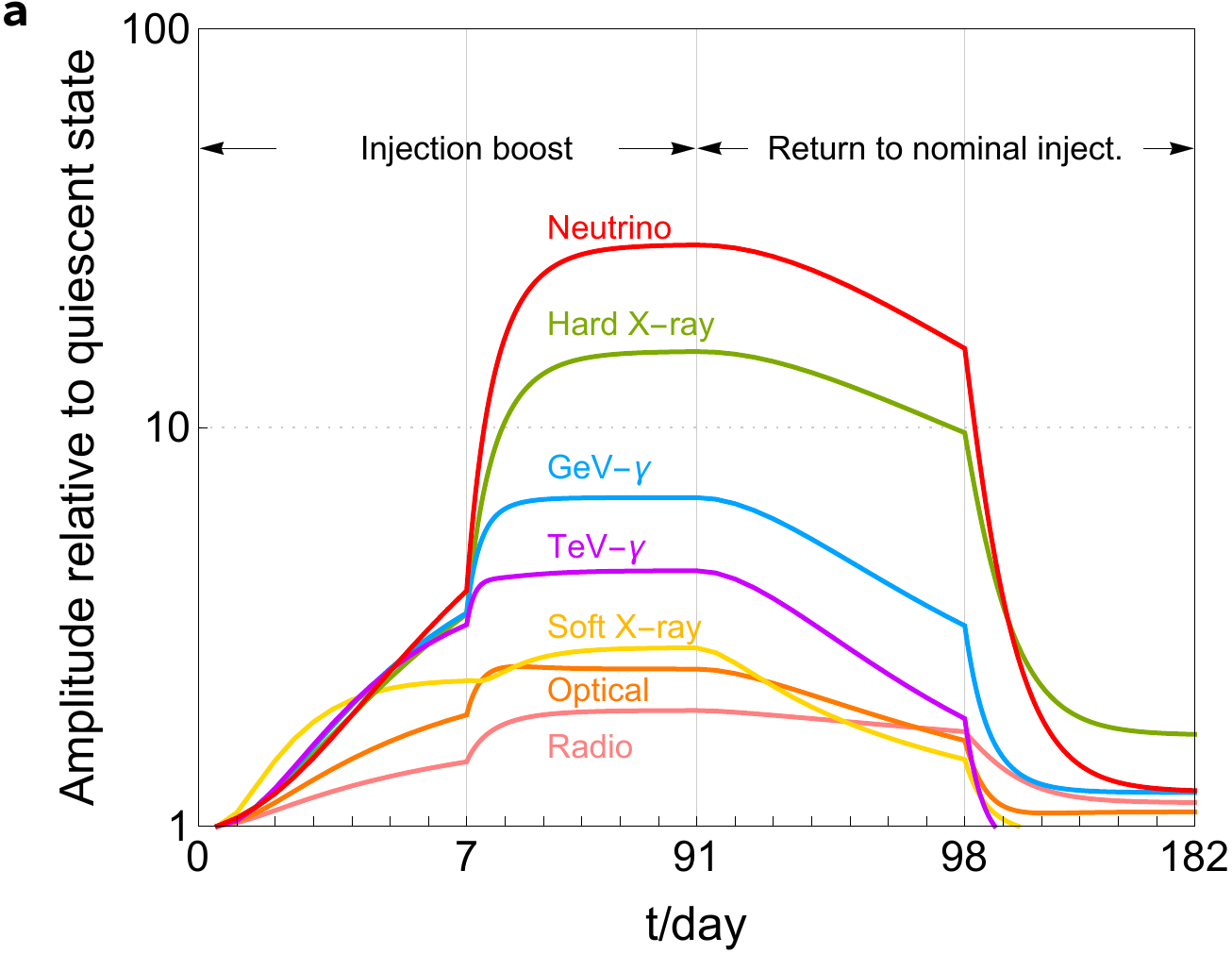}
\includegraphics[width=.463\columnwidth]{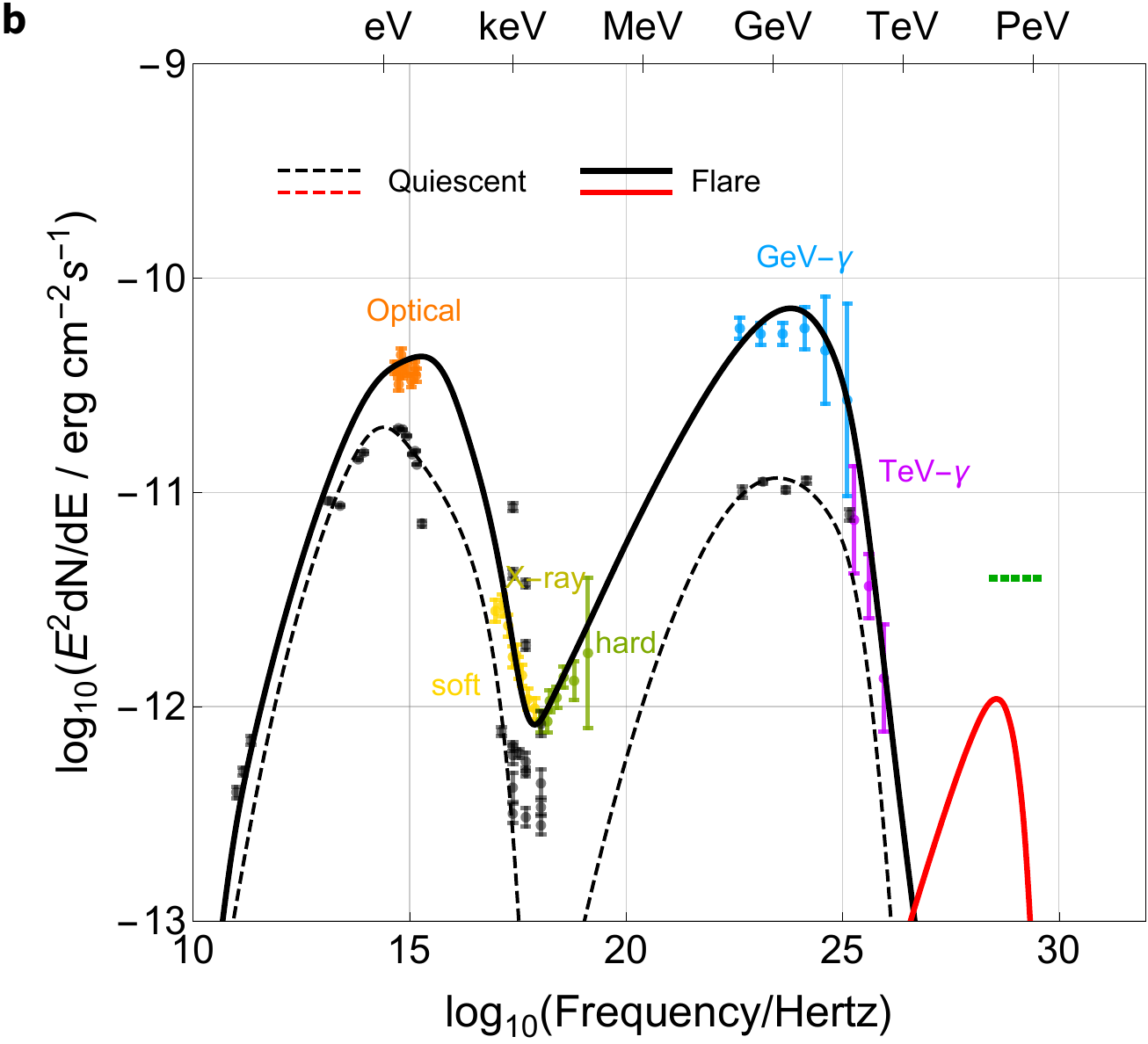}
\end{center}
\caption{\label{fig:TimeDep} Time response of the neutrino and electromagnetic fluxes (left panel) and the SED (right panel) during the electromagnetic flare for a one zone model (2017 neutrino event).  Figure taken from \Ref~\cite{Gao:2018mnu} (Supplementary Materials). }
\end{figure}

It is nevertheless instructive to look at the time-dependence of the one-zone model, as the conclusions for more sophisticated models are similar~\cite{Gao:2018mnu}. We show in \figu{TimeDep} the time response of the SED from the previous one-zone model in different wavelength bands in the left panel, as well as the SED in the quiescent (dashed curve) and electromagnetic flaring state (solid curves) in comparison to respective data in the right panel. It is interesting that the optical peak increases by about a factor of 2.5 during the flare, and the gamma-ray peak by about a factor of $2.5^2 \simeq 6$. This ratio is characteristic for an SSC-dominated model: The synchrotron luminosity scales with the increasing injection of primary electrons, while the GeV gamma-rays from inverse Compton scattering scale with the primary injection $\times$ target density (synchroton photons), \ie, with the square of the primary injection. The neutrino luminosity scales with electron injection (determining the target density) $\times$ proton injection. During the electromagnetic flare, this product needs to be boosted at least by about a factor of ten from the duty cycle of the source~\cite{Murase:2018iyl} -- otherwise, there would be no reason to expect the neutrino during the electromagnetic flare instead of the quiescent period. In certain models (\eg, proton inverse Compton, pion peak) the second hump will scale with electron $\times$ proton injection as well. The increase by a factor of ten (or higher) would contradict the observed factor of six, and therefore these models can be independently ruled out from the time-response of the system. 

In order to address all challenges including the source energetics, more sophisticated source models or geomtries have been proposed, which can be summarized as ``multi-zone models'' with more than one radiation zone. Examples include the formation of a compact core~\cite{Gao:2018mnu}, a layer-sheath geometry of the jet~\cite{Ahnen:2018mvi}, external radiation fields boosted into the jet frame~\cite{Keivani:2018rnh}, and a separate dense region leading to gas interactions~\cite{Liu:2018utd}. These models have  in common that the freedom of the model is increased by an additional radiation zone described by additional parameters. While some models give a plausible explanation for the time-response of the system (such as the compact core model), this aspect is frequently neglected. For example, if the second hump is dominated by external Compton scattering, the description of the correlated activity between the optical/UV and gamma-ray energy ranges will require some conspiracy between electron injection and the external field geometry. Future detections of similar flares will help to refine the models and to identify the most promising alternatives.

\section{The 2014-15 neutrino flare}

The 2014-15 neutrino flare is qualitatively different from the 2017 event since there was no significant electromagnetic activity during that time and a relatively large number of neutrinos has been observed. One challenge is that the neutrino production comes together with a similar amount of energy in photons, as it is evident from the $\Delta$-resonance simplification of photo-hadronic interactions:
\begin{equation}
 p + \gamma \rightarrow \left\{ \begin{array}{ll} 
 n + \pi^+ & \text{in 1/3 of all cases} \\
 p + \pi^0 & \text{in 2/3 of all cases} \\
 \end{array} \right. \, , 
\end{equation}
where the $\pi^0$ decays into two photons, and the $\pi^+$ into three neutrinos and one positron. Thus, the level of the expected neutrino flux determines the level of the injected gamma-ray flux. There are three possibilities discussed in the literature how to ``hide'' the injected energy in photons~\cite{Rodrigues:2018tku,Murase:2018iyl,Wang:2018zln}:
\begin{enumerate}
 \item Since electromagnetic data are sparse during the neutrino flare, the energy may be re-processed through various processes into energy ranges without data, such as the sub-eV or MeV ranges. This implies that electromagnetic monitoring across the whole spectrum will be needed.
 \item 
  The photons may leave the source and be dumped into the EBL, which means that it may not be possible to associate them to a source. This requires relatively low radiation densities in the source (in order to allow the photons to leave), and leads to similar issues with the energetics of the engine, as outlined earlier.
 \item
  The gamma-rays may be absorbed or scattered in an opaque region, such as dust, gas, or radiation around the source. This requires an unusual amount of additional model ingredients. 
\end{enumerate}

\begin{figure}[t]
\begin{center}
\includegraphics[width=0.9\textwidth]{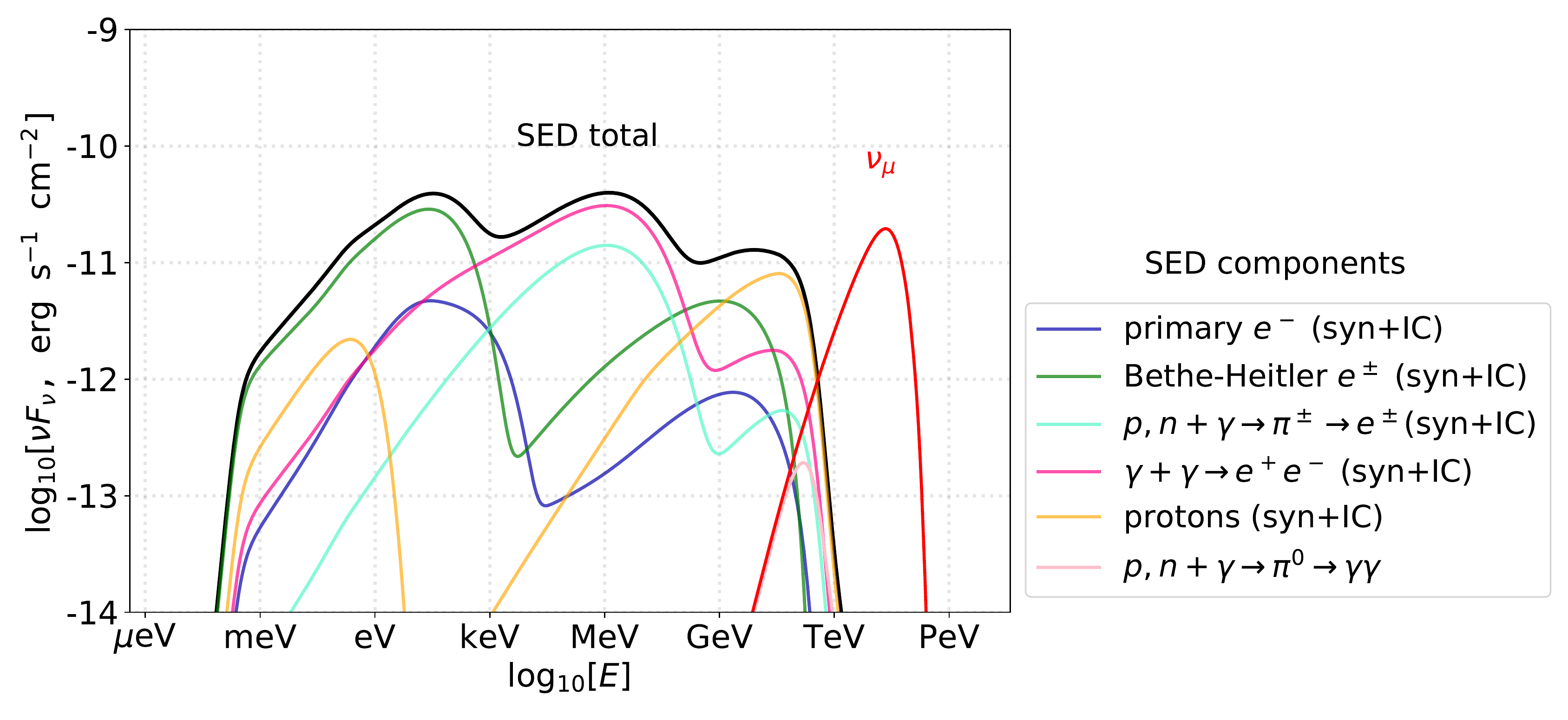} 
\end{center}
\caption{\label{fig:sed_contr} SED and neutrino flux for the one zone model for the 2014-15 neutrino flare (yielding two neutrino events) from \Ref~\cite{Rodrigues:2018tku} (compare to Fig. 1, left panel, red curve therein). Here the different contributions to the SED are highlighted, as indicated in the legend. Note that in this non-trivial model the radiation from primary electrons is always sub-dominant. Figure taken from \Ref~\cite{ThesisRodrigues}.
}
\end{figure}

Especially option~1 requires a critical discussion if that is indeed feasible. An example for a one-zone model is shown in \figu{sed_contr}, where the individual contributions to the SED are highlighted. In this case (predicting two neutrino events at 100~TeV energies), about 80\% of the gamma-ray energy injected beyond TeV energies is actually absorbed in the EBL and 20\% re-processed in the source, which shows up as an additional hump at MeV energies. It is noteworthy that in this example the primary electron emission (synchrotron and inverse Compton, dark blue) dominates nowhere in the SED, which means that this model will not be captured by any conventional classification scheme. Such a model can only be found if all relevant radiation processes are treated self-consistenly; consequently, the result looks different from {\em ad hoc} assumptions of the injected gamma-ray spectrum, \eg~\cite{Halzen:2018iak}.

\begin{figure}[t]
\begin{center}
\begin{tabular}{cc}
\includegraphics[width=0.52\textwidth]{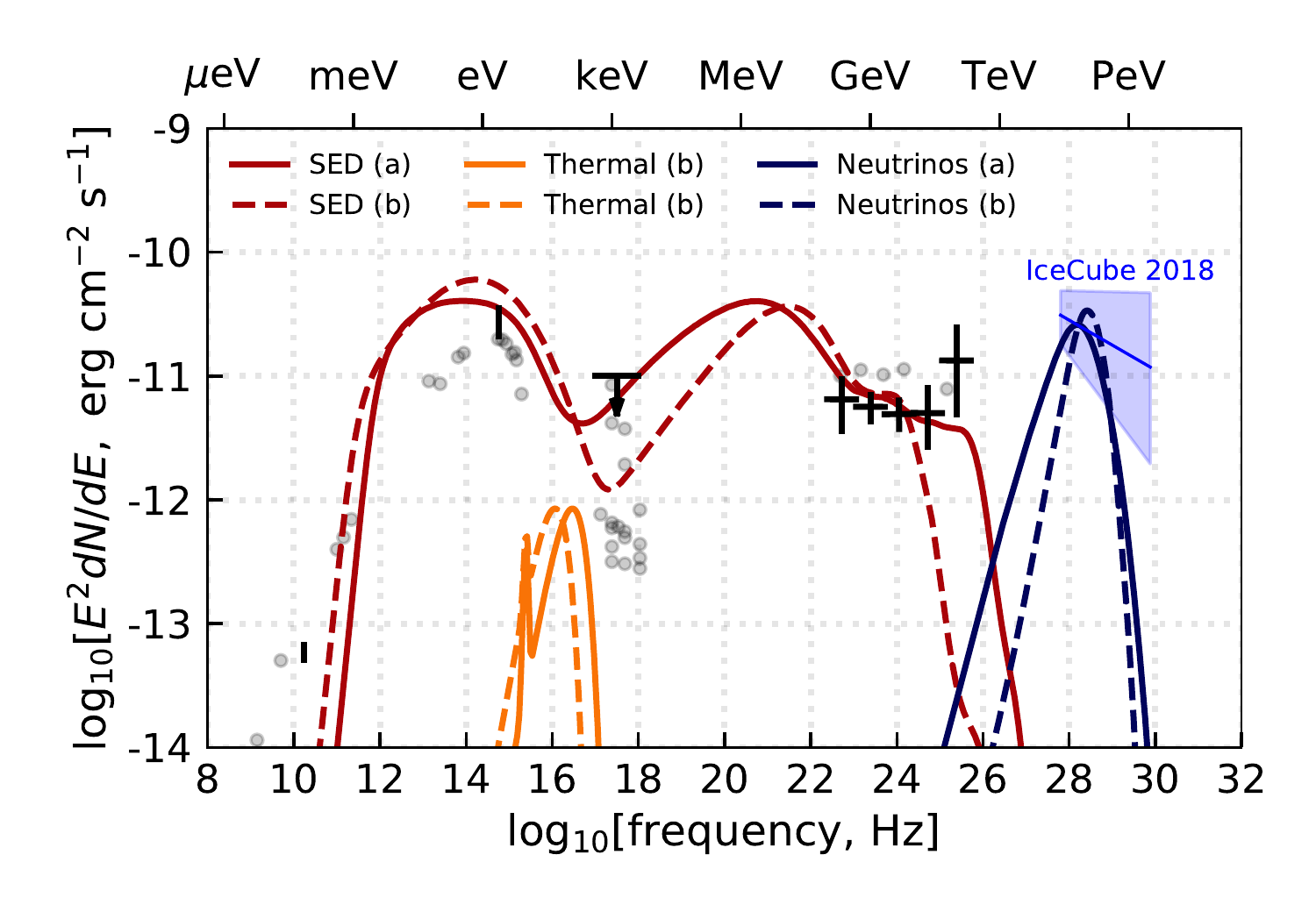} & 
\raisebox{0.3cm}{\includegraphics[width=0.44\textwidth]{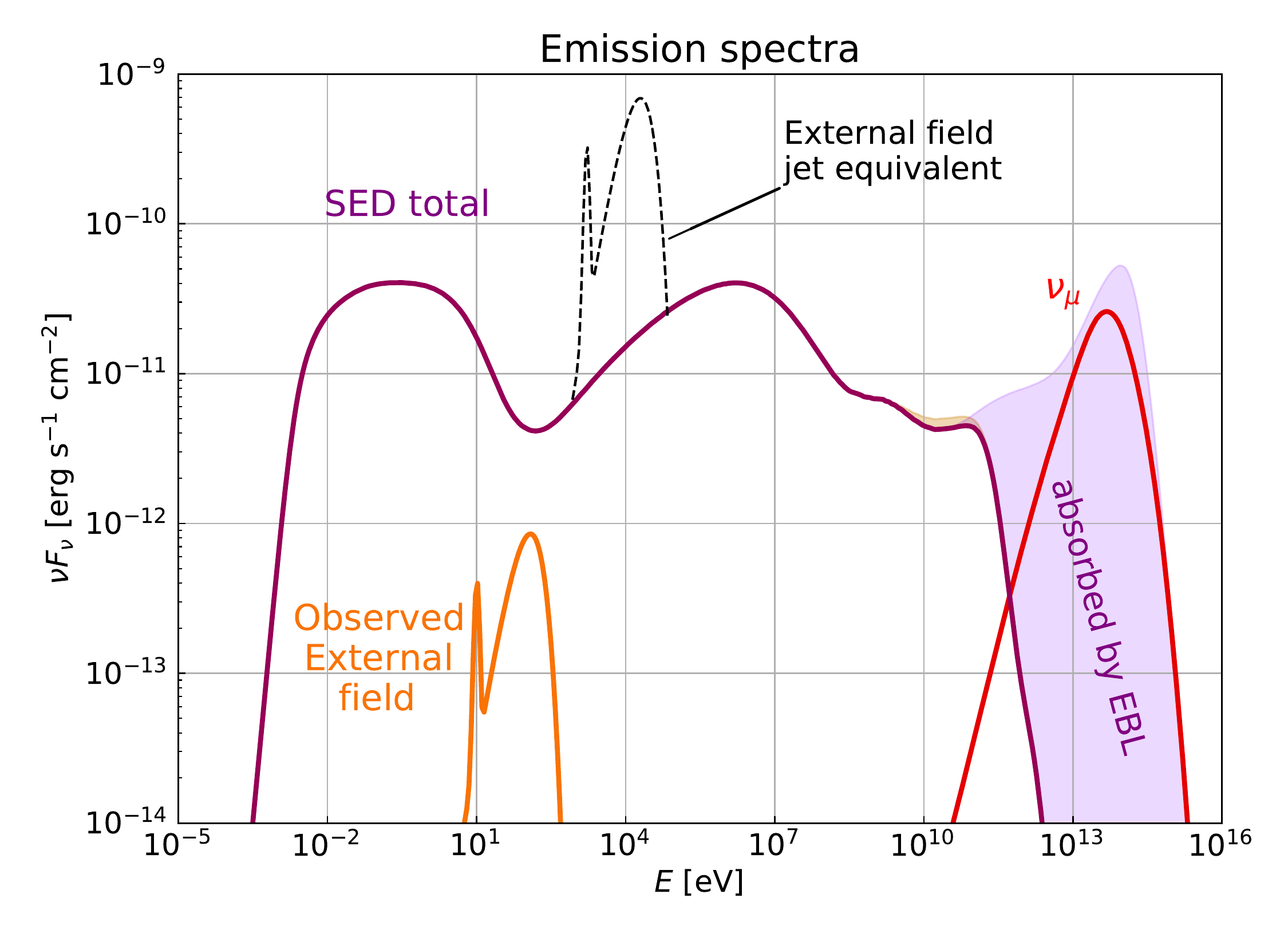}} \\
\end{tabular}
\end{center}
\caption{\label{fig:sed_ext} Left panel: Two examples for an external radiation field model (mostly radiation from accretion disk scattered in BLR) yielding 4.9 and 4.0 neutrino events during the 2014-15 neutrino flare for models a) and b), respectively, as indicated in the legend. Black data points are taken during the neutrino flare, gray data points correspond to archival data taken at other times.  Figure taken from \Ref~\cite{Rodrigues:2018tku}. Right panel: Same as model a) in left panel, with external radiation field seen by observer (orange) and equivalent boosted into blob frame (dashed) -- where it becomes effective for neutrino production. The purple shaded region illustrates the radiation absorbed in the extragalactic background light, the (small) orange shaded region the radiation absorbed in the BLR.  }
\end{figure}

High neutrino production efficiencies can be achieved by external radiation from the disk, back-scattered into the jet in the broad line region (BLR), and from the line emission produced in the BLR in situ. This assumption is supported by the recent hint that TXS 0506+056 may be an FSRQ instead of a BL Lac~\cite{Padovani:2019xcv}, which means that the broad lines are hidden in the SED. We show such an example from \Ref~\cite{Rodrigues:2018tku} in \figu{sed_ext}, where in the right panel the impact of the external radiation field (disk radiation and broad lines) is illustrated both in the observer (orange) and the jet (black) frames. While the radiation is sub-dominant in the observer's frame, the relativistic motion of the blob leads to an extremely intense radiation field in the blob frame. The left panel (solid curves) shows the resulting SED compared to data during the flare (black), with similar features compared to the one-zone model shown earlier: the MeV peak and a large amound of EBL attenuation (illustrated in the right panel). However, here the expected neutrino event numbers between four and five events are close to observations. Futhermore, the prediction of the spectral neutrino shape could be tested in the IceCube analysis. Note that there is a trend of gamma-ray absorption at the highest energies in the blob and in the BLR kicking in at about 10~GeV if the neutrino production is efficient (see also \cite{Reimer:2018vvw}), but the details depend on the energies of the external field and the model assumptions, such as the radiation density and size of the BLR. A gamma-ray hardening, as argued for in \cite{Padovani:2018acg}, can be achieved in a compact core model~\cite{Rodrigues:2018tku} or a jet-cloud model~\cite{Wang:2018zln}.

\section{Discussion and conclusions}

Simple one-zone photo-hadronic models may describe the 2017 neutrino event of TXS 0506+056 only if one accepts that either the physical luminosity exceeded the Eddington luminosity significantly, or that the neutrino peak energy was higher than measured. More complicated source geometries can be used to address this issue; however, the time-response of the system points towards an SSC-dominated model with a sub-leading hadronic contribution which can be best tested in X-rays and VHE gamma-rays. The sparser archival data of the 2014-15 neutrino flare allow for some freedom to deposit the large amound of electromagnetic energy coming with the neutrino production; multi-wavelength monitoring across the whole electromagnetic spectrum will be needed in the future to constrain more exotic possibilities -- which can produce up to five neutrino events. A re-analysis of IceCube data could reveal how much the observed event rate depends on the spectrum assumed. 

A common interpretation of the 2014-15 and 2017 observations remains difficult, although one-zone, compact core, external radiation field or jet-cloud models may be promising approaches which have been used to at least model the two neutrino periods independently. However, the lacking electromagnetic activity in the 2014-15 flare is suspicious, and may raise concerns about its statistical significance. Further observations with electromagnetic follow-ups will be needed for solid conclusions.

{\bf Acknowledgments}. WW  would like to thank the participants of the PAX 2019 workshop for discussions on the subject.

\providecommand{\href}[2]{#2}\begingroup\raggedright\endgroup


\end{document}